\documentclass[11pt]{article}

\usepackage[margin=1in]{geometry}

\usepackage{amsmath,amssymb,amsfonts}
\usepackage{graphicx}
\usepackage{booktabs}
\usepackage{hyperref}
\usepackage{xcolor}
\usepackage[numbers]{natbib}
\usepackage[ruled]{algorithm2e}
\usepackage{caption}
\usepackage{subcaption}

\SetAlFnt{\small}
\SetAlCapFnt{\small}
\SetAlCapNameFnt{\small}
\SetAlCapHSkip{0pt}

\providecommand{\Description}[1]{}

\begin{document}

% Title
\title{Spherical Hermite Maps}

\author{
  Mohamed Abouagour \\
  Indiana University \\
  Bloomington, Indiana, USA \\
  \texttt{moabouag@iu.edu}
  \and
  Eleftherios Garyfallidis \\
  Indiana University \\
  Bloomington, Indiana, USA \\
  \texttt{elef@iu.edu}
}

\date{}

\maketitle

\begin{abstract}
Spherical functions appear throughout computer graphics---from spherical harmonic lighting and precomputed radiance transfer to view-dependent appearance in neural radiance fields, and from mesh level-of-detail impostors to procedural planet rendering.
Efficient evaluation of such functions is critical for real-time applications, yet existing approaches face a quality--performance trade-off: bilinear lookup table (LUT) sampling is fast but produces faceting and discontinuous gradients, while bicubic filtering requires 16 texture samples per query.
Both can provide derivatives in principle, but in practice most implementations use finite differences for normals, requiring additional samples and introducing noise.

This paper presents \emph{Spherical Hermite Maps}, a derivative-augmented LUT representation that resolves this trade-off.
By storing function values alongside scaled partial derivatives at each texel of a padded cubemap, bicubic-Hermite reconstruction is enabled from only four texture samples (a $2\times 2$ footprint) while providing continuous ($C^0$) gradients from the same samples---no additional texture fetches required.
The key insight is that Hermite interpolation naturally reconstructs smooth derivatives as a byproduct of value reconstruction, making surface normals effectively free in terms of memory bandwidth.

In controlled experiments, Spherical Hermite Maps improve PSNR by 8--41\,dB over bilinear interpolation and are competitive with 16-tap bicubic quality at one-quarter the texture fetch cost.
Compared to ``fast bicubic'' (4 bilinear fetches), Hermite achieves higher reconstruction quality while providing continuous gradients at the same sampling cost.
The analytic normals reduce mean angular error by 9--13\% on complex surfaces while yielding stable specular highlights under camera motion.
Three application domains demonstrate versatility: (1) spherical harmonic glyph visualization for scientific data, (2) radial depth-map impostors for mesh level-of-detail, and (3) procedural planet/asteroid rendering with spherical heightfields.
\end{abstract}


\begin{figure*}[t]
  \centering
  \includegraphics[width=\textwidth]{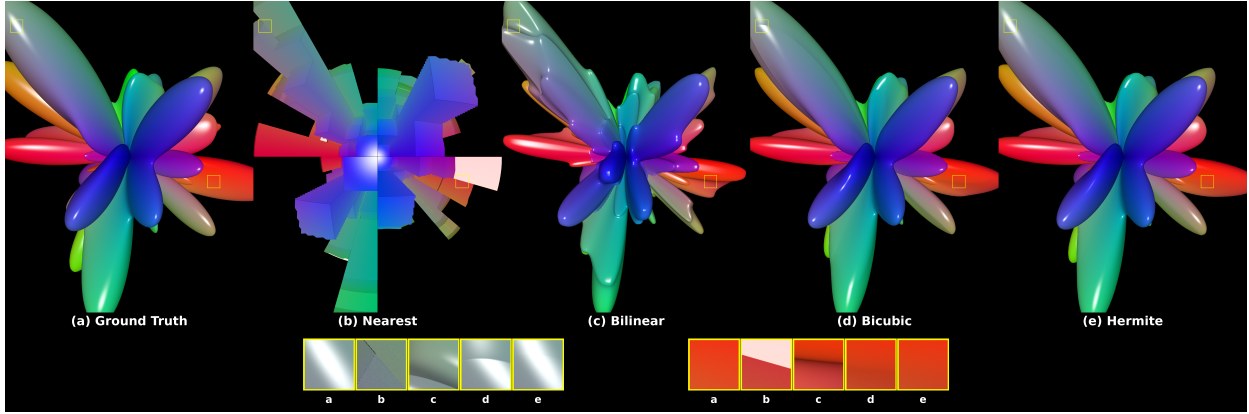}
  \caption{Interpolation at LUT face resolution $N=10$. From left to right: ground truth (direct SH evaluation), nearest-neighbor lookup, bilinear reconstruction with finite-difference normals, 16-tap bicubic reconstruction with finite-difference normals, and Spherical Hermite Maps with analytic normals. Insets zoom a challenging silhouette/highlight region: bilinear exhibits faceting, finite differences introduce gradient noise, and bicubic retains residual error; Hermite interpolation remains faithful to the ground truth at $4\times$ lower sampling cost than 16-tap bicubic.}
  \label{fig:teaser}
\end{figure*}

\section{Introduction}

Spherical functions pervade computer graphics and scientific visualization.
In real-time rendering, spherical harmonics encode view-dependent appearance for precomputed radiance transfer and environment lighting~\citep{sloan2002prt,green2003sh}.
In neural scene representations, they parameterize view-dependent color in radiance fields~\citep{mildenhall2020nerf,kerbl2023gaussian}.
In level-of-detail systems, radial depth maps encode mesh geometry as star-shaped surfaces for impostor rendering.
In procedural content generation, spherical heightfields define asteroid and planetary terrain.
Across these domains, the core computational pattern is the same: evaluate a spherical function at many directions per pixel, often under heavy overdraw, and compute surface normals for shading.

A natural representation for such surfaces is the \emph{star-shaped radial surface}: $x(\omega) = c + R(\omega)\omega$ where $R: S^2 \to \mathbb{R}^+$ defines radius as a function of direction.
This representation includes SH-encoded appearance glyphs, radial depth-map mesh impostors, and procedural planet/asteroid geometry.
Rendering requires both evaluating $R(\omega)$ for ray intersection and computing the spherical gradient $\nabla_{S^2} R$ for surface normals; jointly achieving high-quality reconstruction and stable normals at low cost is a key bottleneck in practice.

A standard acceleration is to precompute samples of the radial function on an angular grid---a per-object lookup table (LUT).
Bilinear sampling is fast but produces visible faceting and unstable specular highlights.
True bicubic filtering improves smoothness but requires 16 texture samples per query, making the shader bandwidth-bound.

Both methods \emph{can} provide derivatives in principle: bilinear interpolation yields gradients that vary within each cell but are discontinuous at cell boundaries, and shader-based bicubic can differentiate its polynomial analytically.
However, \emph{in practice}, most implementations rely on finite differences for normals, which introduces noise and causes highlights to ``swim'' under camera motion.
Even when analytic bicubic derivatives are computed, the 16-sample footprint remains, and the derivatives are only $C^0$ continuous across cell boundaries.

\paragraph{Key insight.}
Hermite interpolation, unlike standard polynomial interpolation, reconstructs a smooth function from both values and derivatives at grid points.
Crucially, the same mathematical formulation that uses stored derivatives for reconstruction also yields continuous first derivatives at query points as a byproduct---from the same four texture samples, without additional texture fetches or finite-difference noise.

This paper presents \emph{Spherical Hermite Maps}, a derivative-augmented LUT representation that exploits this insight.
By storing the function value alongside its first and mixed partial derivatives at each texel, bicubic Hermite reconstruction is enabled from only four texture samples (a $2\times 2$ footprint) while providing analytic gradients for shading without any additional LUT queries.
The result is smooth silhouettes, stable specular appearance, and predictable per-fragment cost---properties important for both interactive graphics applications and scientific visualization workflows.

This paper makes the following contributions: A) Introducing \textbf{Spherical Hermite Maps} enabling high fidelity surface reconstruction from a few samples ($2\times2$) footprint.
B) \textbf{Analytic normals from the same samples}. Surface normals from continuous first derivatives via metric-tensor inversion, without additional texture fetches.
C) \textbf{Exact SH derivative baking}. For spherical harmonic functions, derivatives are computed by differentiating the SH basis analytically, eliminating discretization errors.
D) \textbf{Seam-consistent padding}. Gutter texels ensuring seam-consistent values and derivatives across cubemap faces.

\textbf{Results} are presented in 3 example applications including: a) spherical harmonic (SH) glyphs from real diffusion MRI data, b) mesh level-of-detail (LOD) impostors for 3D games, and c) procedural planets for scientific visualization.

%\paragraph{What is new vs.\ prior work.}
Cubic Hermite interpolation is a mathematical technique for reconstructing $C^1$ functions from values and derivatives on grids, and its tensor-product form extends naturally to 2D~\citep{bartels1987splines}.
Derivative information is also commonly stored in textures for shading purposes (e.g., normal maps, derivative maps) and has been explored for smooth interpolation of sampled scalar fields.
However, making Hermite interpolation practical for spherical (cubemap) LUTs requires non-trivial implementation details: seam-consistent cubemap padding/filtering and geometrically correct transport of per-face chart derivatives to spherical gradients for shading~\citep{isidoro2005angularextent,praun2003sphericalparam}.

A key aspect of this work is to turn Hermite interpolation into a practical spherical LUT primitive for GPU rendering.
A padded cubemap layout with seam-consistent values and derivatives across faces is introduced, metric-correct conversion from chart-space derivatives to spherical gradients via the cubemap metric tensor is derived, and it is demonstrated that the same four RGBA fetches used for value reconstruction provide analytic normals from continuous first derivatives with no additional texture bandwidth.
Crucially, bicubic filtering requires a $4\times 4$ footprint (16 samples) when derivatives are not stored; Hermite achieves bicubic-class reconstruction from a $2\times 2$ footprint (4 samples) because derivatives and the mixed derivative are supplied at the corners.
This is distinct from fast bicubic approximations~\citep{gpugems2}, which match the 4-fetch budget but do not provide robust analytic derivatives and therefore fall back to finite differences for shading.

\section{Related Work}

Many graphics pipelines repeatedly evaluate spherical functions (e.g., spherical harmonics for lighting and view-dependent appearance~\citep{sloan2002prt,green2003sh,mildenhall2020nerf,kerbl2023gaussian}), motivating precomputed spherical lookup tables (LUTs) for real-time rendering and visualization~\citep{peeters2009pacificvis,falk2017glyphraycasting}.
For LUT parameterization, cubemaps avoid pole singularities but require careful seam handling for filtering~\citep{isidoro2005angularextent,praun2003sphericalparam}; padded faces are used and padding is extended to derivatives so both reconstruction and normals remain continuous across boundaries.

Higher-order filtering typically relies on shader bicubic reconstruction (16 samples) or fast bicubic approximations using a few bilinear fetches~\citep{gpugems2}.
Hermite interpolation~\citep{bartels1987splines} achieves $C^1$ continuity from a $2\times2$ footprint when per-texel derivatives are available.
Applying it to spherical (cubemap) LUTs requires seam-consistent padding for both values and derivatives, and a geometrically correct conversion from per-face chart derivatives to spherical gradients via the cubemap metric tensor.

Normals for LUT-rendered surfaces are often estimated with finite differences, trading extra samples for noisy and discontinuous gradients; storing normal/gradient maps can improve shading but does not provide higher-order value reconstruction.
In contrast, Spherical Hermite Maps provide bicubic-class reconstruction and analytic gradients from the same four samples.

\section{Method}

Given spherical harmonic coefficients for each glyph, a derivative-augmented LUT is first baked on a padded cubemap.
At runtime, each glyph is rendered as a billboard impostor, the view ray is intersected with the star-shaped surface, and the hit point is shaded using normals reconstructed analytically from Hermite-interpolated derivatives.

\subsection{Surface Model and Billboard Rendering}

A spherical function $r: S^2 \to \mathbb{R}$ at sample center $c$ is visualized as a star-shaped radial surface:
\begin{equation}
x(\omega) = c + R(\omega)\,\omega, \quad \omega \in S^2,
\end{equation}
where $R(\omega) \geq 0$ is the rendered radius and $s > 0$ is a user-specified scale factor.
For nonnegative functions, $R = s \cdot r$; for signed functions, $R = s \cdot |r|$ with appropriate sign tracking for normal computation.
The implicit form is $F(x) = \|x - c\| - R(\omega)$ where $\omega = (x-c)/\|x-c\|$.

Each glyph is rendered as a camera-facing billboard quad sized to enclose the surface (radius $s \cdot \max_\omega r(\omega)$).
For each fragment, a view ray $p(t) = o + t\,d$ is cast from camera origin $o$ along direction $d$ through the quad.
A bounding sphere of radius $R_{\max}$ is first intersected to obtain an entry interval $[t_{\min}, t_{\max}]$, then the surface hit is refined within this bracket using bisection (typically 8--16 steps) followed by Newton--Raphson iteration (2--4 steps) on $F(p(t)) = 0$.
To avoid unnecessary computation for distant glyphs, projected screen-space size is estimated from bounding radius and depth, allocating more refinement iterations to large projected glyphs.

\subsection{Padded Cubemap Parameterization}

Directions are mapped to cubemap faces with local coordinates $(u, v) \in [0,1]^2$ per face.
This yields six regular 2D grids, avoids pole singularities, and aligns with GPU cubemap hardware.

To ensure continuity across face boundaries, each face is padded with a one-texel gutter.
For interior resolution $N \times N$, each face is stored as $(N+2) \times (N+2)$; interior indices $\{1, \ldots, N\}$ correspond to the original samples.
Crucially, gutter texels are filled by evaluating the spherical function at directions corresponding to their texel centers---not by extrapolation or reflection---so that cross-face interpolation samples correct function values.
Derivatives are treated similarly: derivative channels are baked for every stored texel (interior and gutter) by differentiating the sampled function in the per-face chart coordinates $(u,v)$, rather than copying derivatives across seams.

\subsection{Spherical Hermite Maps}

Standard bicubic filtering achieves $C^1$ continuity but requires a $4 \times 4$ texel footprint (16 samples).
Hermite interpolation achieves the same continuity from a $2 \times 2$ footprint when both values and derivatives are available.
This is exploited by storing four quantities per texel:
\begin{equation}
\bigl(r,\; r_u h,\; r_v h,\; r_{uv} h^2\bigr),
\label{eq:hermite_storage}
\end{equation}
where $h = 1/N$ is the texel spacing and derivatives are pre-scaled to simplify shader arithmetic.

\paragraph{Derivative computation.}
Derivatives are computed in the chart coordinates $(u,v)$ on each face.
In the general case, $r(\omega(u,v))$ is evaluated at texel centers on a padded grid and $(r_u, r_v, r_{uv})$ are estimated using finite-difference stencils in $(u,v)$; to obtain derivatives at gutter texels without sampling ``outside'' the stored texture, a small additional bake-time margin (or one-sided stencils on the outer ring) is used.
For spherical harmonics, derivatives can alternatively be computed analytically in spherical coordinates and converted to $(u,v)$ derivatives via the chain rule.

\begin{algorithm}[t]
\DontPrintSemicolon
\caption{Bake-time Spherical Hermite Map (per face)}
\KwIn{spherical function $r(\omega)$; interior face resolution $N$}
\KwOut{padded $(N+2)\times(N+2)$ RGBA face texture}
$h \leftarrow 1/N$\;
\ForEach{stored texel center $(u_k,v_k)$ (including gutter)}{
$r_k \leftarrow r\bigl(\omega(u_k,v_k)\bigr)$\;
}
Compute chart derivatives $(r_u,r_v,r_{uv})$ on the padded grid via finite differences in $(u,v)$ (using a bake-time margin or one-sided stencils at the outer ring)\;
For SH, compute analytic derivatives and convert to $(u,v)$ via the chain rule.\;
Store per texel as RGBA: $\bigl(r_k,\; (r_u h)_k,\; (r_v h)_k,\; (r_{uv} h^2)_k\bigr)$\;
\end{algorithm}

\paragraph{Analytic SH derivatives.}
For spherical harmonic functions, \emph{exact analytic derivatives} can be computed directly from SH coefficients by differentiating the basis functions.
Since each real SH basis function $Y_l^m(\theta, \phi)$ has a known closed-form derivative with respect to spherical coordinates, the derivative of any SH expansion $r(\omega) = \sum_{l,m} c_l^m Y_l^m(\omega)$ is simply $\partial r / \partial \theta = \sum_{l,m} c_l^m \, \partial Y_l^m / \partial \theta$ (and similarly for $\phi$).
This eliminates finite-difference discretization error entirely at bake time, producing derivatives that are exact to floating-point precision.
For high-order SH (large $L$), where the function contains fine angular detail, analytic baking preserves accuracy that central differences would otherwise smooth away.
The implementation supports both modes: central differences for generality (procedural functions, mesh distance fields) and analytic SH differentiation for maximum accuracy when coefficients are available.
At each sampled direction $\omega$, analytic derivatives are converted to face-coordinate derivatives via the chain rule, e.g., $r_u = \nabla_{S^2} r \cdot \partial\omega/\partial u$ and $r_v = \nabla_{S^2} r \cdot \partial\omega/\partial v$.

\paragraph{Derivative accuracy and limitations.}
When using central differences, stored derivatives are analytic with respect to the \emph{Hermite interpolant}, not the underlying continuous function.
This approximation introduces discretization error proportional to $h^2$ (the grid spacing squared), which decreases rapidly as LUT resolution increases.
For undersampled high-frequency content, derivative estimates may be noisy---though no worse than runtime finite differences, and with the advantage of smooth $C^1$ interpolation across cells.
When exact derivatives are required, analytic SH baking (described above) avoids finite-difference discretization error.
% (up to floating-point roundoff).

\paragraph{Reconstruction.}
Given local coordinates $(s,t) \in [0,1]^2$ within a cell, reconstruction proceeds using the tensor-product cubic Hermite basis.
The 1D Hermite basis functions are:
\begin{align}
H_0(s) &= 2s^3 - 3s^2 + 1, & H_1(s) &= -2s^3 + 3s^2, \\
H_0^d(s) &= s^3 - 2s^2 + s, & H_1^d(s) &= s^3 - s^2.
\end{align}
The 2D reconstruction is:
\begin{multline}
r(s,t) = \sum_{i,j \in \{0,1\}} \bigl[ H_i H_j \, r_{ij} + H_i^d H_j \, (r_u h)_{ij} \\
+ H_i H_j^d \, (r_v h)_{ij} + H_i^d H_j^d \, (r_{uv} h^2)_{ij} \bigr],
\end{multline}
where subscripts $ij$ index the four cell corners.
This uses all 16 quantities from the four corners (4 channels $\times$ 4 texels), yielding bicubic-class reconstruction.

\paragraph{Derivative extraction.}
Chart derivatives $(r_u, r_v) = (\partial r / \partial u, \partial r / \partial v)$ are the partial derivatives with respect to the local cubemap face coordinates, distinct from the spherical gradient $\nabla_{S^2} r$ which lives in the tangent plane of the sphere.
These chart derivatives are obtained by differentiating the Hermite basis:
\begin{equation}
r_u(s,t) = \frac{1}{h} \sum_{i,j} \bigl[ H_i' H_j \, r_{ij} + (H_i^d)' H_j \, (r_u h)_{ij} + \cdots \bigr].
\end{equation}
These derivatives emerge from the same four texture samples used for value reconstruction---no additional LUT queries are needed.

\subsection{Analytic Normals from Chart Derivatives}

For a star-shaped surface $x(\omega) = c + R(\omega)\omega$, the outward normal involves the spherical gradient $\nabla_{S^2} R$.
This is obtained by transporting chart derivatives through the cubemap Jacobian using metric tensor inversion---a key step that ensures geometric correctness.

Let $\omega(u,v)$ be the normalized direction for face coordinates $(u,v)$ and let $e_1 = \partial\omega/\partial u$ and $e_2 = \partial\omega/\partial v$ be the (unnormalized) tangent vectors.
These span the tangent plane at $\omega$ but are generally neither orthogonal nor unit length.

In practice, $e_1$ and $e_2$ are computed by differentiating the normalized cubemap direction.
Let $\tilde{\omega}(u,v)$ be the unnormalized direction (affine in $(u,v)$ on each face) and $\omega = \tilde{\omega}/\|\tilde{\omega}\|$.
Differentiating the normalization gives:
\begin{equation}
\frac{\partial \omega}{\partial \xi}
= \frac{(I_3 - \omega\omega^\top)}{\|\tilde{\omega}\|}\,\frac{\partial \tilde{\omega}}{\partial \xi},
\quad \xi \in \{u, v\},
\end{equation}
and $\partial\tilde{\omega}/\partial u$, $\partial\tilde{\omega}/\partial v$ are constant per face.

The tangent-space gradient $\nabla_{S^2} R = \alpha e_1 + \beta e_2$ satisfies:
\begin{equation}
\nabla_{S^2} R \cdot e_1 = R_u, \quad \nabla_{S^2} R \cdot e_2 = R_v.
\end{equation}
This yields a 2$\times$2 linear system involving the metric tensor $G$:
\begin{equation}
\begin{bmatrix} g_{11} & g_{12} \\ g_{12} & g_{22} \end{bmatrix}
\begin{bmatrix} \alpha \\ \beta \end{bmatrix}
= \begin{bmatrix} R_u \\ R_v \end{bmatrix},
\end{equation}
where $g_{11} = e_1 \cdot e_1$, $g_{22} = e_2 \cdot e_2$, $g_{12} = e_1 \cdot e_2$.
Solving via Cramer's rule:
\begin{equation}
\alpha = \frac{g_{22} R_u - g_{12} R_v}{\det G}, \quad
\beta = \frac{g_{11} R_v - g_{12} R_u}{\det G}.
\end{equation}
Near cube corners where $\det G$ may be small, clamping before inversion ensures numerical stability.

The surface normal is then:
\begin{equation}
n = \frac{R(\omega)\,\omega - \nabla_{S^2} R}{\|R(\omega)\,\omega - \nabla_{S^2} R\|}.
\end{equation}
For signed functions with $R = s|r|$, the chain rule is applied: $\nabla_{S^2} R = s \cdot \mathrm{sign}(r) \cdot \nabla_{S^2} r$.

\paragraph{The ``free normals'' property.}
Normals are obtained from the same Hermite evaluation used for value reconstruction: chart derivatives $(R_u, R_v)$ are produced by differentiating the Hermite basis, so shading does not require finite-difference neighbor samples.

\section{Experiments}

This section evaluates Spherical Hermite Maps along three axes: reconstruction accuracy, visual quality, and sampling efficiency.
The experiments compare three table-based reconstruction methods---bilinear, 16-tap bicubic, and Hermite---against direct spherical harmonic evaluation as ground truth.
All methods share the same padded cubemap parameterization and differ only in reconstruction.

\subsection{Experimental Setup}

\paragraph{Synthetic benchmarks.}
A representative SH glyph is rendered at degree $L=8$ (fixed seed) with direct SH evaluation as ground truth.
To avoid conflating value reconstruction and normal estimation, two matched metrics are reported:
(i) \emph{value-only} error by sampling $r(\omega)$ over many directions on $S^2$, and
(ii) \emph{shaded-image} error by computing PSNR over rendered images under fixed lighting.
For each cubemap face resolution $N \in \{8, 12, 16, 24, 32, 48, 64\}$, the same coefficient field and camera setup are evaluated.

\paragraph{Visual comparisons.}
For qualitative assessment, synthetic SH glyphs and real diffusion MRI orientation distribution functions are rendered under controlled lighting.
Silhouette smoothness, specular highlight stability, and seam artifacts at cubemap face boundaries are examined.

\paragraph{Implementation.}
The renderer is implemented in WGSL and accessed via pygfx~\citep{pygfx2024} and FURY~\citep{fury2020}, which provide WebGPU bindings.
Precomputation is performed on the CPU during scene initialization; runtime rendering uses only GPU-resident textures.

\subsection{Reconstruction Accuracy}

\begin{figure}[t]
  \centering
  \includegraphics[width=0.7\columnwidth]{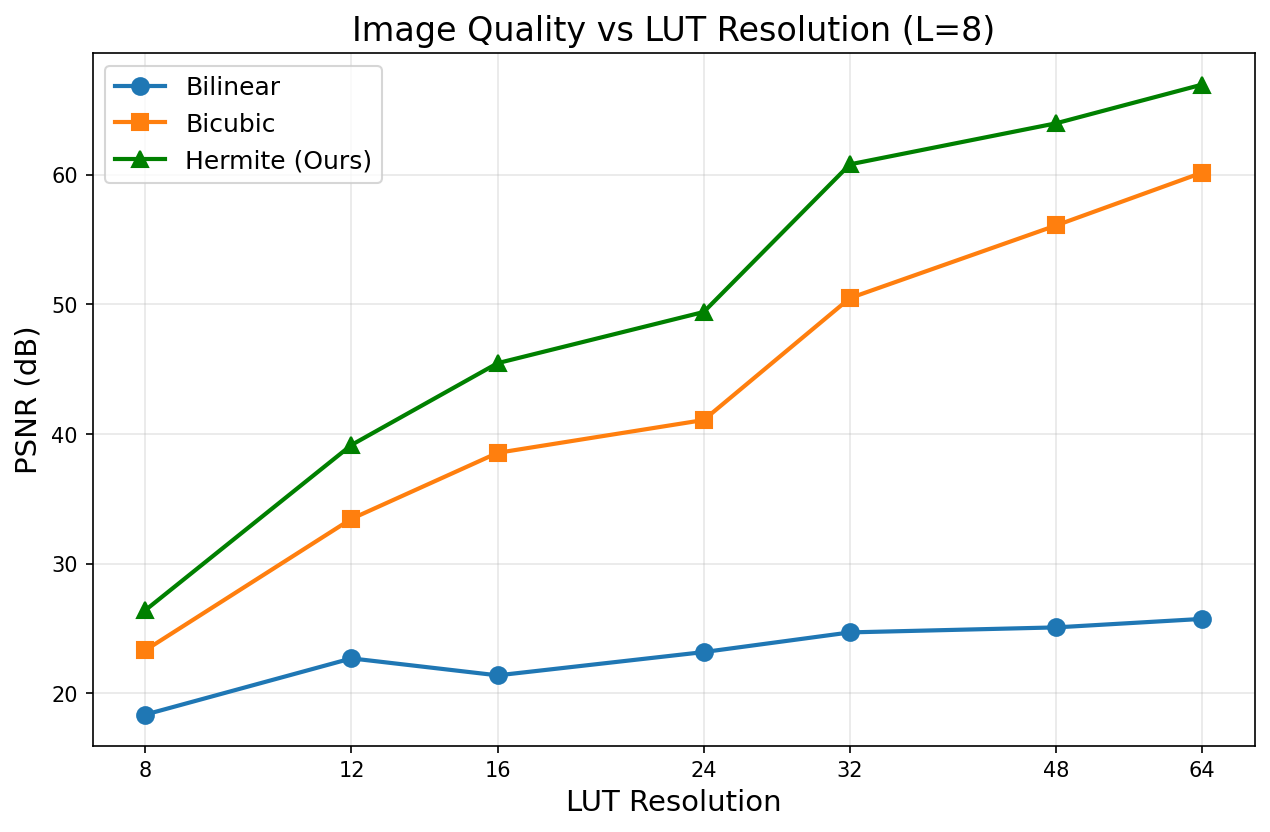}
\caption{Image-based reconstruction accuracy versus LUT resolution for a representative $L=8$ spherical-harmonic glyph. Plot shows PSNR (higher is better) of LUT-based rendering compared to direct SH evaluation. Hermite consistently improves over bilinear across resolutions and exceeds 16-tap bicubic quality while using only a $2\times2$ (four-sample) footprint.}
  \Description{A line plot showing PSNR versus LUT resolution for bilinear, bicubic, and Hermite reconstruction methods.}
  \label{fig:psnr_comparison}
\end{figure}

Figure~\ref{fig:psnr_comparison} summarizes the \emph{shaded-image} PSNR benchmark (direct SH rendering as ground truth).
Hermite reconstruction significantly outperforms bilinear sampling, with improvements ranging from +8.0\,dB at $N=8$ to +36.1\,dB at $N=32$ and +41.3\,dB at $N=64$.
Compared to 16-tap bicubic filtering, Hermite achieves higher PSNR (+3.1\,dB at $N=8$, +10.3\,dB at $N=32$) while using only 4 texture samples instead of 16.
Table~\ref{tab:psnr_value} reports the corresponding \emph{value-only} PSNR for $r(\omega)$ sampled over random directions; it shows the same method ordering, confirming that Hermite's advantage is not driven by normal estimation.

\begin{table}[t]
  \centering
  \small
\caption{Value-only PSNR (dB) for the sampled radius function $r(\omega)$ vs.\ direct SH evaluation (100K uniform random directions on $S^2$). Bicubic uses an interpolating Catmull--Rom kernel. Absolute dB values are not directly comparable to image-based PSNR, but the ranking is consistent.}

  \label{tab:psnr_value}
  \setlength{\tabcolsep}{6pt}
  \begin{tabular}{rcccc}
    \toprule
    $N$ & Bilinear & Bicubic & Hermite & $\Delta$(H--Bic) \\
    \midrule
    8   & 23.1 & 29.1 & \textbf{32.2} & +3.1 \\
    16  & 35.5 & 51.5 & \textbf{62.7} & +11.2 \\
    32  & 48.0 & 73.0 & \textbf{93.1} & +20.1 \\
    \bottomrule
  \end{tabular}
\end{table}
In the shaded-image benchmark, bilinear plateaus around 25\,dB regardless of resolution because its reconstruction order limits accuracy even with more samples.
Both bicubic and Hermite continue improving with resolution, with Hermite approaching 67\,dB at $N=64$---near the limit of single-precision floating point.

\subsection{Visual Quality}

Figure~\ref{fig:teaser} shows a representative close-up comparison at fixed face resolution ($N=10$).
Bilinear sampling exhibits visible faceting along silhouettes.
Bicubic filtering (16 taps) improves smoothness but retains residual error at this resolution.
Hermite reconstruction produces smooth silhouettes from a $2\times 2$ footprint; the corresponding highlight behavior is evaluated next.

\subsection{Shading Quality and Highlight Stability}

% Beyond geometric reconstruction, shading quality depends critically on normal accuracy.
% Figure~\ref{fig:shading} compares highlight behavior across methods.
% Finite-difference normals (used by bilinear, and often paired with bicubic baselines in practice) produce noisy gradients that manifest as irregular highlight shapes and ``swimming'' artifacts during camera motion.
% Hermite's analytic normals yield smooth, predictable highlights that remain stable under animation.

% \begin{figure*}[t]
%   \centering
%   \includegraphics[width=\textwidth]{figures/fig_shading_comparison_professional.png}
%   \caption{Multi-glyph shading comparison at $N=10$. From left to right: ground truth, bilinear + finite-difference normals, bicubic + finite-difference normals, and Hermite with analytic normals. Insets highlight visible artifacts caused by discontinuous or noisy gradients (silhouette waviness and unstable specular structure); Hermite produces smooth silhouettes and stable highlights from the same four-sample reconstruction.}
%   \Description{Grid of SH glyphs rendered with Hermite interpolation showing smooth shading quality.}
%   \label{fig:shading}
% \end{figure*}

\begin{figure*}[t]
  \centering
  \includegraphics[width=\textwidth]{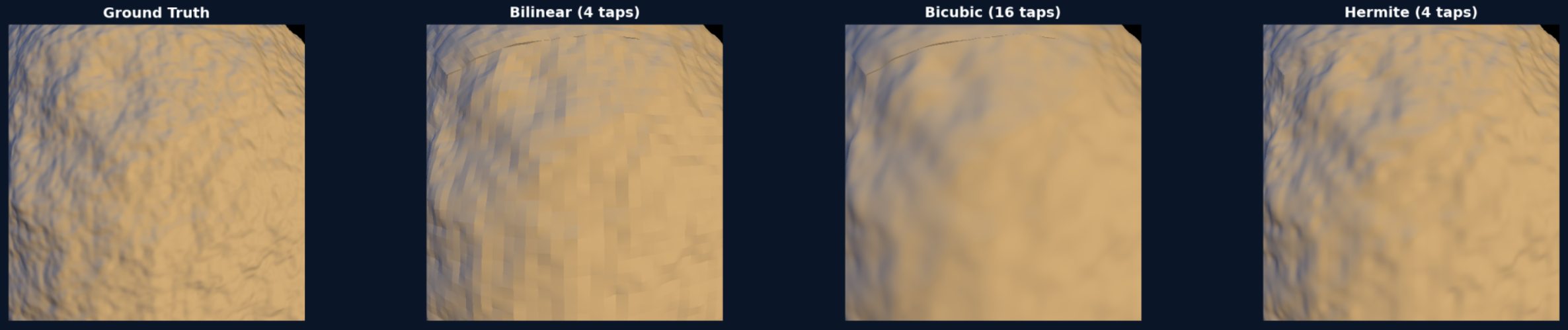}
  \caption{Procedural planet terrain comparison at $N=48$. Shaded surface. From left to right: ground truth, bilinear (4 taps), bicubic (16 taps), and Hermite (4 taps). Hermite achieves the lowest normal error (7.28°) while outperforming  bicubic's sample efficiency advantage over the 16-tap baseline.}
  \Description{Four-column comparison of planet terrain showing ground truth, bilinear, bicubic, and Hermite methods with shaded views and normal maps.}
  \label{fig:planet_detail}
\end{figure*}

Beyond geometric reconstruction, shading quality depends critically on normal accuracy.
Figure~\ref{fig:planet_detail} compares shaded terrain and normal maps across methods on a procedural planet close-up.
Finite-difference normals (used by bilinear and bicubic) produce noisy gradients visible in both the shaded surface and normal map.
Hermite's analytic normals yield smoother gradients (7.28° mean error vs.\ 7.82°--7.97° for alternatives) and more stable highlights.

\subsection{Sampling Efficiency and Bandwidth Analysis}

Table~\ref{tab:cost_extended} summarizes the per-query cost in terms of texture instructions (individual fetch operations) and scalar reads (a proxy for memory bandwidth).
A distinction is made between \emph{texture instructions} (GPU texture unit operations) and \emph{scalars read}.

Hermite uses 4 texture instructions (one RGBA fetch per corner of a $2\times 2$ cell), compared to 16 for point-sampled bicubic.
Although hardware bilinear filtering requires only 1 texture instruction for values, shaded rendering typically adds at least 4 offset samples for finite-difference normals (5 total).
Hermite has access to additional derivative information, so the relevant comparison is quality per texture instruction (and at equal scalar storage), rather than interpolation order alone.

\paragraph{Bandwidth considerations.}
Hermite stores 4 scalars per texel (RGBA) rather than 1 for a value-only texture.
Each query therefore reads $4 \times 4 = 16$ scalar values, which remains competitive with bilinear+FD shaded rendering at $5 \times 4 = 20$ scalar values (assuming the FD stencil overlaps with the original sample).
More importantly, Hermite's 4-instruction footprint is friendly to texture cache locality and GPU scheduling, whereas the 16-instruction bicubic footprint increases cache pressure on complex scenes.

\paragraph{Comparison with fast bicubic approximations.}
A common optimization approximates bicubic filtering using 4 hardware-bilinear fetches instead of 16 point samples~\citep{gpugems2}.
This ``fast bicubic'' approach is an important baseline; Table~\ref{tab:cost_extended} includes this comparison.

\begin{table}[t]
  \centering
  \small
  \caption{Extended cost comparison including fast bicubic. ``Tex ops'' stands for texture instructions; ``Scalars'' stands for scalar values read for a single query (multiply by 4 bytes for FP32).}
  \label{tab:cost_extended}
  \setlength{\tabcolsep}{4pt}
  \begin{tabular}{lcccc}
    \toprule
    Method & Tex ops & Scalars & $C^1$ value & continuous $\nabla$ \\
    \midrule
    Bilinear (HW) & 1 & 4 & No & No \\
    Bilinear + FD & 5 & 20 & No & No \\
    Bicubic (16 pt) & 16 & 16 & Yes & Yes$^\dagger$ \\
    Fast Bicubic (4 bilin) & 4 & 16 & Yes & No$^*$ \\
    Fast Bicubic + FD & 8 & 32 & Yes & No \\
    \textbf{Hermite} & \textbf{4} & \textbf{16} & \textbf{Yes} & \textbf{Yes} \\
    \bottomrule
  \end{tabular}
  
  \vspace{2pt}
  {\footnotesize $^\dagger$Gradients can be obtained by differentiating the bicubic reconstruction (extra ALU, same samples); this is reported as ``Bicubic + analytic $\nabla$'' in Tables~\ref{tab:asteroid_normals}--\ref{tab:planet_normals}.}
  
  {\footnotesize $^*$Fast bicubic uses weighted bilinear; differentiating it is non-trivial and rarely used.}
\end{table}

Fast bicubic matches the 4-instruction budget but does not provide readily usable gradients, and typically falls back to finite differences for shading (8 texture ops, 32 scalars).
Hermite provides both values and continuous gradients from the same 4 fetches.

\paragraph{Equal-storage comparison.}
\label{sec:equal-storage}
Since Hermite stores 4 values per texel, a fair comparison considers what value-only methods achieve with the same memory budget.
At equal storage (4$N^2$ floats per face), Hermite at resolution $N$ competes against bilinear at resolution $2N$.

\begin{figure*}[p]
  \centering
  \includegraphics[width=\textwidth]{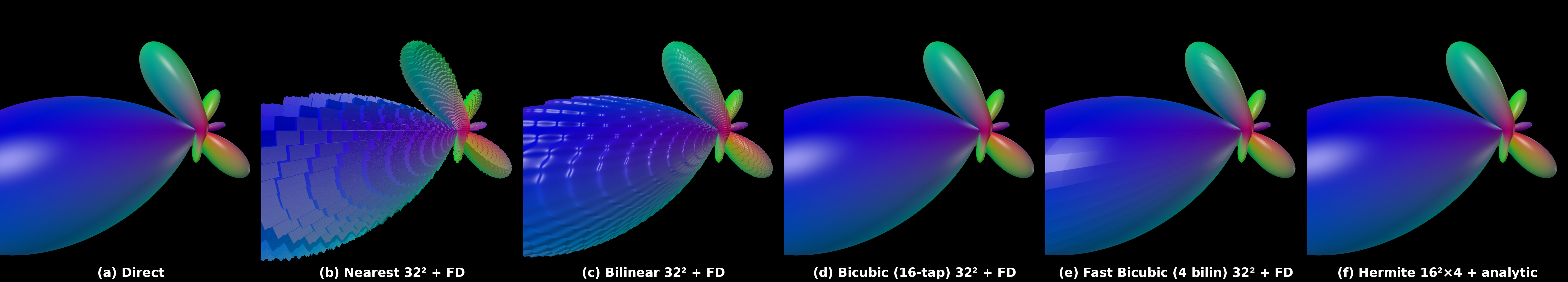}
  \caption{Equal-storage comparison (shaded SH glyph). At matched scalar storage per face (32$^2\times$1ch vs 16$^2\times$4ch), Hermite achieves high quality results using only a 2$\times$2 footprint and generates a slightly higher image PSNR than 16-tap bicubic (+1.2\,dB), while also providing analytic derivatives for shading. Fast bicubic shows visible banding under specular lighting due to finite-difference normals, showing that 4 fetches alone are insufficient without analytic gradients.}
  \Description{Equal-storage comparison showing SH glyph rendered with value-only LUT at 32x32 resolution versus Hermite at 16x16 with analytic gradients.}
  \label{fig:equal_storage}
\end{figure*}

Figure~\ref{fig:equal_storage} compares value-only sampling at 32$^2$ against Hermite at 16$^2$ with four channels (value + derivatives), which matches scalar storage per face.
In this representative view, Hermite slightly exceeds 16-tap bicubic in image PSNR (+1.2\,dB) while using only 4 fetches and providing analytic derivatives for shading.

In the shaded-image benchmark, bilinear at $N=16$ achieves 21.4\,dB PSNR, while Hermite at $N=8$ achieves 26.4\,dB---a +5.0\,dB advantage for Hermite at the same storage.
At higher resolutions, Hermite at $N=16$ (45.5\,dB) versus bilinear at $N=32$ (24.7\,dB) shows a +20.8\,dB advantage.

The key insight is that Hermite's 4$\times$ storage overhead is offset by its 4$\times$ reduction in texture instructions \emph{and} its analytic derivatives.
When rendering shaded glyphs at equal storage, bilinear typically uses 5 texture instructions (1 value + 4 finite-difference normals), while Hermite provides value and normals in 4.
Thus Hermite achieves better PSNR, better normals, and 20\% fewer texture operations at equal storage, and is preferred under shaded rendering at equal storage.

\subsection{Runtime Performance}

To validate that theoretical sampling efficiency translates to practical performance gains, rendering throughput is benchmarked on two GPUs: an integrated Intel Iris Xe (laptop) and a discrete NVIDIA RTX.
Table~\ref{tab:runtime} reports frame rates at face resolution $N=8$.

\begin{table}[t]
  \centering
  \small
  \caption{Runtime performance across two GPU classes. Integrated: 100K glyphs on Intel Iris Xe. Discrete: 1M glyphs on NVIDIA RTX. Hermite consistently achieves $1.8\times$ speedup over bicubic.}
  \label{tab:runtime}
  \setlength{\tabcolsep}{4pt}
  \begin{tabular}{lccccc}
    \toprule
    & \multicolumn{2}{c}{Intel Iris Xe (100K)} & \multicolumn{2}{c}{NVIDIA RTX (1M)} & \\
    Method & FPS & Ratio & FPS & Ratio & continuous $\nabla$ \\
    \midrule
    Nearest & 14.6 & 2.9$\times$ & 32 & 3.6$\times$ & No \\
    Bilinear & 10.0 & 2.0$\times$ & 23 & 2.6$\times$ & No \\
    \textbf{Hermite} & \textbf{9.0} & \textbf{1.8}$\times$ & \textbf{16} & \textbf{1.8}$\times$ & \textbf{Yes} \\
    Bicubic & 5.1 & 1.0$\times$ & 9 & 1.0$\times$ & No$^*$ \\
    \bottomrule
  \end{tabular}
  
  \vspace{2pt}
  {\footnotesize $^*$Bicubic uses FD normals for fair comparison. Ratio = speedup vs bicubic.}
      \vspace{-1em}

\end{table}

As a reference, direct per-fragment SH evaluation without precomputation averaged 0.35~FPS on Intel Iris Xe at 100K glyphs (more than an order of magnitude slower than any precomputed-table method), so we omit it from Table~\ref{tab:runtime}.

Across both GPUs, Hermite achieves $1.8\times$ higher frame rate than 16-tap bicubic while providing comparable reconstruction and analytic gradients for shading.
On integrated graphics (Intel Iris Xe), Hermite is only 10\% slower than bilinear (9.0 vs 10.0 FPS).
On discrete graphics (RTX 2000 Ada) with 10$\times$ more glyphs, Hermite remains 30\% slower than bilinear (16 vs 23 FPS) but still delivers the same $1.8\times$ speedup over bicubic.
This consistent speedup ratio across hardware classes validates that Hermite's efficiency advantage is fundamental rather than platform-specific.

\subsection{Mipmap Derivative Handling}

A practical consideration for production use is mip-chain construction.
When building lower-resolution mip levels for Hermite cubemaps, derivatives must be handled carefully.
Table~\ref{tab:lod} compares two approaches: filtering values and recomputing derivatives at each mip level (``Consistent'') versus box-filtering all RGBA channels (``Naive'', as standard hardware mipmapping would do).

\begin{table}[t]
  \centering
  \small
  \caption{LOD quality: derivative-consistent vs.\ naive mip generation.
           Normal error (degrees) measured against filtered ground truth.
           Naive mipmapping causes up to 5$\times$ worse normals at the first coarse LOD.}
  \label{tab:lod}
  \setlength{\tabcolsep}{3pt}
  \begin{tabular}{rcc|cc|cc}
    \toprule
    & & \multicolumn{2}{c|}{Correct} & \multicolumn{2}{c|}{Naive} & \\
    Level & Res & Mean & 95th & Mean & 95th & Ratio \\
    \midrule
    0 & $32^2$ & 0.4° & 0.8° & 0.4° & 0.8° & 1.0$\times$ \\
    1 & $16^2$ & 1.3° & 3.4° & 7.3° & 17.1° & \textbf{5.5$\times$} \\
    2 & $8^2$ & 5.4° & 13.2° & 15.6° & 36.9° & 2.9$\times$ \\
    3 & $4^2$ & 21.0° & 63.0° & 26.8° & 70.8° & 1.3$\times$ \\
    \bottomrule
  \end{tabular}
  
  % \vspace{2pt}
  % 
  {\footnotesize SH degree $L=8$. ``Correct'' recomputes derivatives per level; 
   ``Naive'' box-filters all RGBA channels. Mip chains built offline.}
   
\end{table}

At mip level~1 (16²), naive filtering produces 5.5$\times$ reduced normal accuracy than derivative-consistent generation (7.3° vs.\ 1.3° mean angular error).
The 95th percentile shows even larger gaps: 17.1° vs.\ 3.4°.
This occurs because box-filtered derivatives no longer represent the gradients of the filtered signal and break the scaled derivative interpretation in Eq.~\ref{eq:hermite_storage}.

To preserve Eq.~\ref{eq:hermite_storage} under filtering, we build mip levels by filtering the scalar field and then recomputing derivative channels from the filtered values using finite differences in $(u,v)$ (central differences where available).
Derivative-consistent mip generation is therefore essential for preserving normal quality under LOD.

\subsection{Application: Diffusion MRI Visualization}

\begin{figure*}[p]
  \centering
  \includegraphics[width=\textwidth]{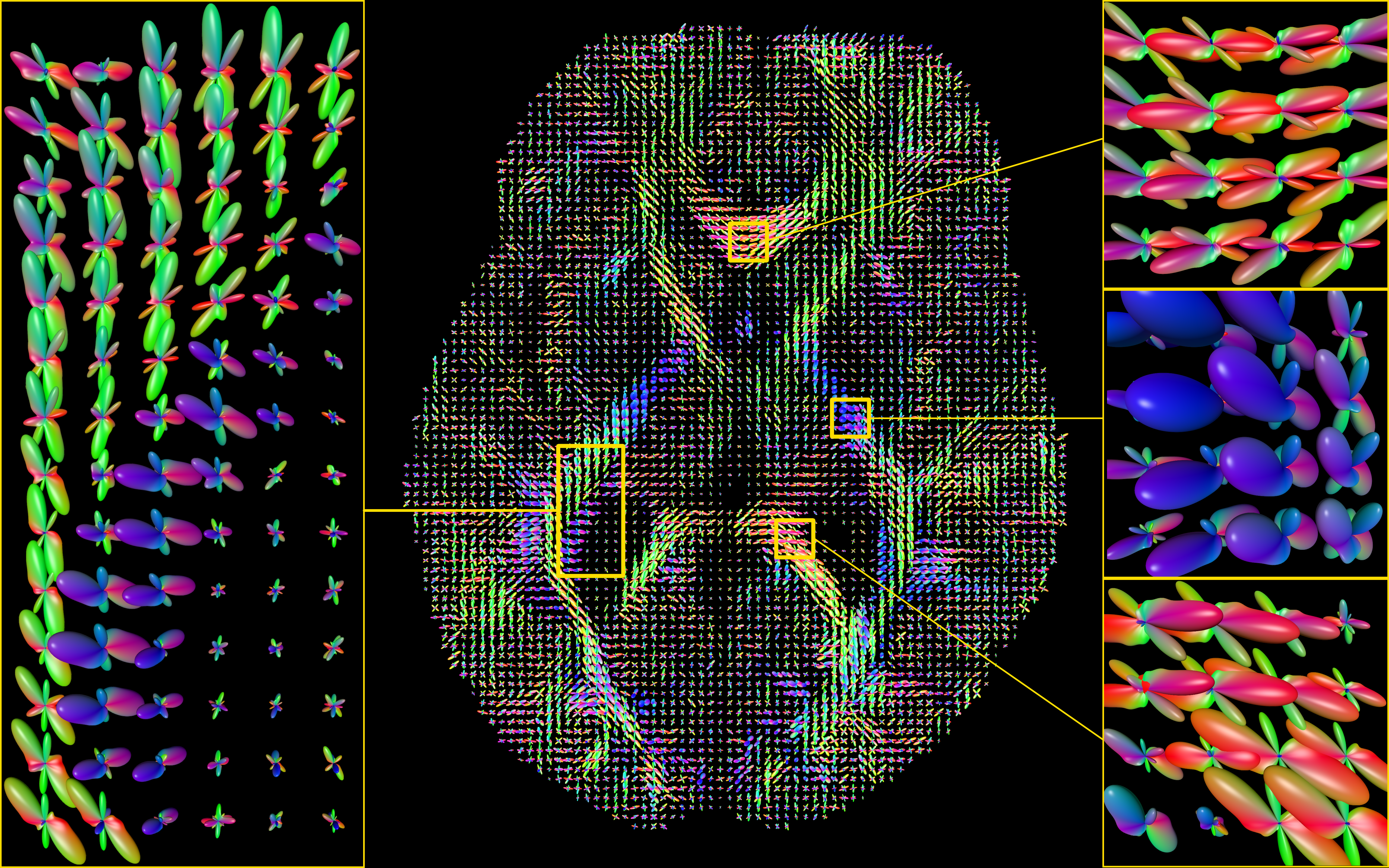}
  \caption{Diffusion MRI orientation distribution function (ODF) glyphs from a human brain axial slice rendered at 4K resolution using our proposed Spherical Hermite Maps. Color encodes fiber orientation. The analytic normals produced by our reconstruction yield high quality silhouettes and stable specular highlights, supporting reliable visual interpretation of the underlying white-matter pathways.}
  \Description{A 2D slice of diffusion MRI ODF glyphs from a human brain, showing complex fiber orientations with smooth shading.}
  \label{fig:dmri}
\end{figure*}

To validate practical applicability, Spherical Hermite Maps are applied to real diffusion MRI data.
Figure~\ref{fig:dmri} shows orientation distribution function (ODF) glyphs from a human brain dataset, rendered at 4K resolution. Data and SH fits from DIPY \citep{garyfallidis2014dipy}.
The smooth silhouettes and stable highlights enable reliable visual interpretation of fiber orientations, which is critical for clinical and research applications.
Interactive frame rates ($>$30 FPS) are achieved on commodity hardware, enabling real-time exploration of volumetric ODF fields. 

\subsection{Application: Mesh LOD Impostors}

A practical application of Spherical Hermite Maps is radial depth-map impostors for mesh level-of-detail.
Given a complex mesh, the radial depth function $R(\omega) = \max_{t} \{t : c + t\omega \in \text{mesh}\}$ is precomputed, where $c$ is the mesh centroid.
This encodes the mesh as a star-shaped surface that can be rendered via ray casting against a cubemap texture, providing a view-independent LOD representation.

These impostors are bandwidth-sensitive because normals are typically estimated with finite differences (4 additional samples per shading query), introducing noise and highlight instability.
Hermite reconstruction provides analytic chart derivatives from the same 4 samples used for depth reconstruction, enabling stable shading without additional texture fetches.

\begin{figure*}[p]
  \centering
  \includegraphics[width=\textwidth]{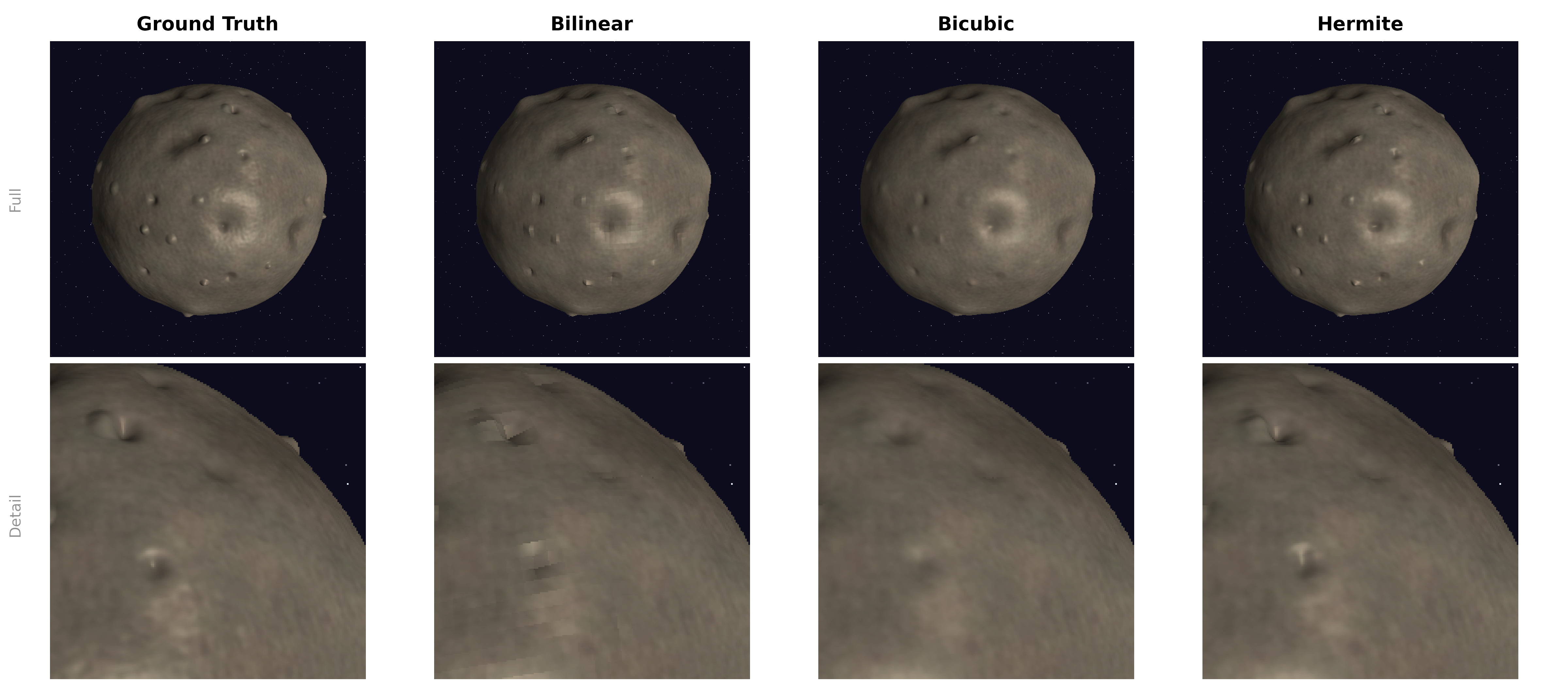}
  \caption{Asteroid mesh impostor reconstructed from a $48\times48$ LUT per cubemap face. Top row: full view; bottom row: close-up. Columns show ground truth, bilinear, bicubic, and Hermite. Insets emphasize crater rims and boulder textures where finite-difference normals are most visibly noisy. Hermite achieves 13\% lower mean normal error and +1.2\,dB PSNR using only four texture samples (vs.\ 8--20 for alternatives).}
  \Description{Four-column comparison of asteroid impostor rendering showing ground truth, bilinear, bicubic, and Hermite methods with shaded and detail views.}
  \label{fig:asteroid_impostor}
\end{figure*}

Figure~\ref{fig:asteroid_impostor} demonstrates this on a complex asteroid model with 69 craters (6 large, 18 medium, 45 small), 8 tectonic ridges, 30 boulders, and 5-octave fractal noise.
At face resolution $48\times48$:

\begin{table}[ht]
  \centering
  \small
  \caption{Quantitative metrics on complex asteroid impostor (48$\times$48 LUT).}
  \label{tab:asteroid_normals}
  \setlength{\tabcolsep}{4pt}
  \begin{tabular}{lccccc}
    \toprule
    Method & Samples & PSNR & Mean Err & 95th \%ile & Improv \\
    \midrule
    Bilinear + FD & 4+4=8 & 73.1 dB & 4.13° & 19.3° & --- \\
    Bicubic + FD & 16+4=20 & 72.3 dB & 4.32° & 20.3° & -5\% \\
    Bicubic + analytic $\nabla$ & 16 & 72.7 dB & 4.19° & 19.3° & -1\% \\
    \textbf{Hermite} & \textbf{4} & \textbf{74.3 dB} & \textbf{3.60°} & \textbf{16.7°} & \textbf{+13\%} \\
    \bottomrule
  \end{tabular}
\end{table}

Hermite achieves 13\% lower mean normal error than bilinear while using \emph{half} the texture samples (4 vs 8) and 1.2~dB higher PSNR.
Compared to bicubic, Hermite uses 5$\times$ fewer samples than the finite-difference baseline (4 vs 20) while achieving 17\% better normal accuracy.
Even when bicubic normals are computed analytically by differentiating the bicubic kernel (no extra samples), Hermite still achieves lower normal error while using 4$\times$ fewer samples (4 vs 16).
The error reduction is most pronounced on high-frequency features like crater rims and boulder edges, where finite-difference noise is worst.

\subsection{Application: Procedural Planet Rendering}

Spherical Hermite Maps also enable efficient rendering of procedural planets and asteroids defined by spherical heightfields.
A common approach for planet rendering is to define terrain as $R(\omega) = R_0 + h(\omega)$ where $h(\omega)$ is a procedural noise function (Perlin, Simplex, or multi-octave fBm).

For real-time applications, evaluating procedural noise per-pixel is expensive.
Baking the terrain to a Hermite map amortizes noise evaluation and provides analytic normals for terrain shading.
This is particularly valuable for distant planets where noise detail exceeds pixel resolution, as the map naturally provides filtered values.

\begin{figure*}[p]
  \centering
  \includegraphics[width=\textwidth]{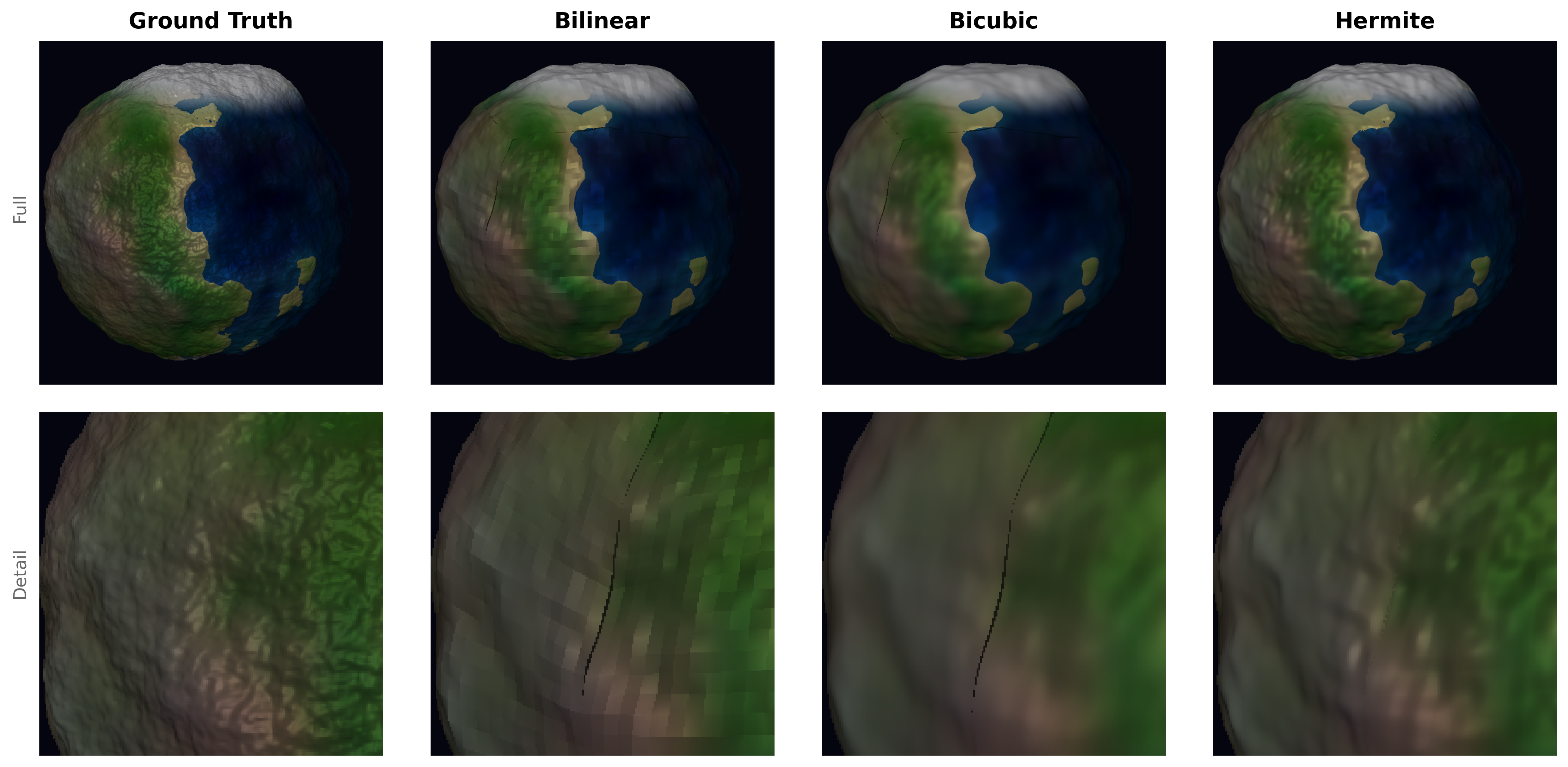}
  \caption{Procedural planet terrain baked into a $48\times48$ LUT per cubemap face. Insets highlight multi-scale terrain regions where interpolation differences are most apparent. Hermite reduces mean normal error by 9\% while requiring only four texture samples.}
  \Description{Four-column comparison of procedural planet rendering showing ground truth, bilinear, bicubic, and Hermite methods.}
  \label{fig:planet_terrain}
\end{figure*}

Figure~\ref{fig:planet_terrain} shows a procedural planet with multi-scale terrain features: large continental landmasses, mountain ranges, craters, and fine-scale surface variation.
At face resolution $48\times48$:

\begin{table}[ht]
  \centering
  \small
  \caption{Quantitative metrics on procedural planet terrain (48$\times$48 LUT).}
  \label{tab:planet_normals}
  \setlength{\tabcolsep}{4pt}
  \begin{tabular}{lccccc}
    \toprule
    Method & Samples & PSNR & Mean Err & 95th \%ile & Improv \\
    \midrule
    Bilinear + FD & 4+4=8 & 23.6 dB & 7.94° & 16.1° & --- \\
    Bicubic + FD & 16+4=20 & 23.6 dB & 8.08° & 15.7° & -2\% \\
    Bicubic + analytic $\nabla$ & 16 & 24.1 dB & 7.87° & 15.5° & +1\% \\
    \textbf{Hermite} & \textbf{4} & \textbf{24.8 dB} & \textbf{7.24°} & \textbf{14.8°} & \textbf{+9\%} \\
    \bottomrule
  \end{tabular}
\end{table}

The improvement is more modest than for the asteroid case because planetary terrain has more gradual variations, but Hermite still achieves 9\% lower error and 1.2~dB higher PSNR while avoiding finite-difference sampling.
Analytic bicubic derivatives reduce normal error marginally (+1\% in Table~\ref{tab:planet_normals}) but do not change the 16-tap footprint.
For distant planets rendered at low screen resolution, Hermite achieves this accuracy from a $2\times 2$ footprint (four taps).

\section{Discussion}
\label{sec:discussion}

As shown experimentally Spherical Hermite Maps deliver high-order reconstruction with fewer texture samples than bicubic while providing analytic gradients---a practical alternative for applications requiring spherical function LUTs with accurate shading. Spherical Harmonic Maps are validated across spherical harmonic glyph visualization, mesh LOD impostors, and procedural planet rendering demonstrating wide range applicability.

% \paragraph{When to use Spherical Hermite Maps.}
\paragraph{Advantages.}
This new approach is primarily beneficial when shading quality matters and LUT evaluation dominates per-fragment cost (e.g., dense spherical fields). For sparse glyphs with a few spherical functions, storage of $4\times$ may not be necessary.

\paragraph{Limitations.}
Parameterization. Folded charts (e.g., octahedral maps) introduce derivative discontinuities without special handling at folds.
Very low resolutions. At $N<12$, seam artifacts can appear due to discrete derivative estimates; higher resolutions mitigate this.
Storage overhead. Hermite stores four scalars per texel (value plus three derivatives); for tight memory budgets, higher-resolution value-only LUTs may be preferable.
Mip/LOD generation. Derivatives must be recomputed per mip level; na\"ive box-filtering of all RGBA channels breaks semantics and degrades normals (Table~\ref{tab:lod}).

\section{Conclusion}

This work introduces Spherical Hermite Maps, a derivative-augmented lookup table representation for efficient visualization of complex spherical functions using Hermite polynomials.

By storing function values alongside scaled partial derivatives on a padded cubemap, we enable high fidelity interpolation from a $2\times 2$ texel footprint with analytic surface normals at no additional sampling cost.

Our experiments demonstrate substantial improvements:
a) +8--41\,dB PSNR over bilinear; exceeding 16-tap bicubic quality, b) 9--13\% lower normal angular error on complex surfaces and c) $1.8\times$ speedup over bicubic interpolation in Intel and NVIDIA GPUs.

%Our approach is validated across spherical harmonic glyph visualization, mesh LOD impostors, and procedural planet rendering. Spherical Hermite Maps deliver high-order reconstruction with fewer texture samples than bicubic while providing analytic gradients---a practical alternative for applications requiring spherical function LUTs with accurate shading.

\clearpage
% Figure-only pages (floats only; no section headings or body text).

\begin{figure*}[p]
  \centering
  \includegraphics[width=\textwidth]{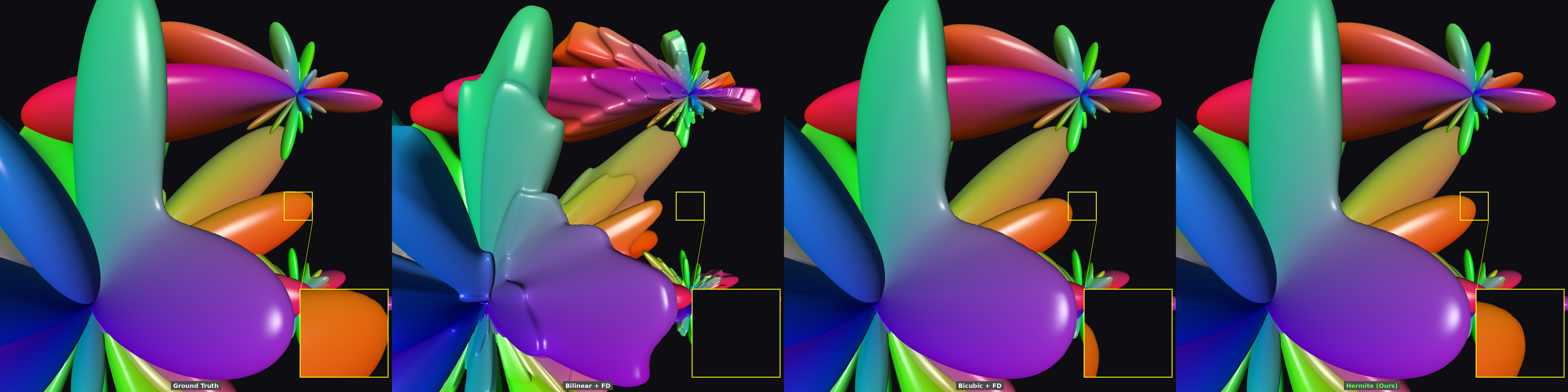}
  \caption{Multi-glyph shading comparison at $N=10$. From left to right: ground truth, bilinear + finite-difference normals, bicubic + finite-difference normals, and Hermite with analytic normals. Insets highlight visible artifacts caused by discontinuous or noisy gradients (silhouette waviness and unstable specular structure); Hermite produces smooth silhouettes and stable highlights from the same four-sample reconstruction.}
  \Description{Grid of SH glyphs rendered with Hermite interpolation showing smooth shading quality.}
  \label{fig:shading}
\end{figure*}

\clearpage

\bibliographystyle{plainnat}
\bibliography{references}

\end{document}